# Clustering Properties of Stars in Simulations of Wind-Driven Star Formation

John Scalo and David Chappell

*Astronomy Department, University of Texas, Austin, TX 78712*

**ABSTRACT**

Several recent observational studies have shown that the clustering of young stars in local star-forming regions, and of Cepheids in the LMC, can be described by a power law two-point correlation function. We show by numerical simulations that the observed range in power law slopes can be accounted for by a model in which stellar winds drive expanding shells that are subjected to nonlinear fluid advection and interactions with other shells, and in which star formation occurs when a threshold shell column density is exceeded. The models predict how the power law slope should depend on the maximum age of the stellar sample and the average star formation rate, although a number of effects preclude a comparison with currently-available data. We also show how stellar migration flattens the power law slope below a scale that depends on the velocity dispersion and age of the sample, an effect which may explain the secondary breaks in the observed correlation functions of some regions at large separations. Problems with using the correlation function as a descriptor of clustering structure for statistically inhomogeneous data sets are discussed.

## 1. INTRODUCTION

A number of recent papers have attempted to quantify the clustering of young stars using the angular two-point correlation function, beginning with Gomez et al. (1993) for the Taurus region, in the hope that this function can constrain theoretical models for star formation. Larson (1995) used the equivalent average surface density of pairs as a function of separation, which we denote $\Sigma_p(\Delta r)$, to show that, when binary survey data were included with the data used by Gomez et al. (1993), $\Sigma_p(\Delta r)$ consisted of two power laws, as had been suggested by Gomez et al. For separations less than about 0.04 pc in Taurus the power law index $\gamma$ is steep, with a value of –2.15, which he identified with the binary star regime. At larger separations the power law index is flatter, with a value of –0.62. Larson suggested that this "clustering regime" reflected the hierarchical, or fractal, spatial distribution of the gas from which the stars formed, and that the break in $\Sigma_p(r)$ at 0.04 pc is a signature of the Jeans length for the gas in cool dense cloud cores.

Simon (1997) extended Larson's results to include Taurus, the $\rho$ Oph core, and the Orion Trapezium cluster. The form of $\Sigma_p(r)$ was similar to that found by Larson, but with evidence for region-to-region variations in $\gamma$ for the clustering regime, with a range –0.3 to –0.7. In addition, Simon showed that the dependence of the break scale on region properties was very difficult to reconcile with an interpretation in terms of the Jeans length, and suggested instead that the break scale simply reflects the average stellar separation. Simon's conclusions were strengthened by Nakajima et al. (1998, hereafter NTHN), who estimated $\Sigma_p(r)$ for young stars in three Orion regions and the $\rho$ Oph, Chamaeleon, Vela, and Lupus star-forming regions. NTHN find a range in $\gamma$ from –0.15, for the region they call Orion OB, to –0.82, for Lupus. Their results also demonstrate that the break scale is about a tenth of the mean nearest-neighbor separation for all the regions, suggesting again that this scale is not an imprint of the Jeans length, which should scale as mean separation to the 3/2 power (assuming star density proportional to density of gas from which they form).

Gomez & Lada (1998) studied the two-point correlation function for optically-selected stars in the $\lambda$ Ori portion of the Orion OB association and the Ori A region, and find power laws with indices between –0.21 for the H$\alpha$ stars in Ori A and –0.49 for the $\lambda$ Ori region. They also found features in the $\lambda$ Ori correlation function that could be interpreted as the sizes and separations of the subclusters. (See Houlahan & Scalo 1990 for an analytical demonstration of this effect.)

Elmegreen & Efremov (1996) studied the spatial distribution of Cepheid variables in the LMC on much larger scales ($\sim$0.1 to 2 Kpc) than the studies of local star-forming regions discussed above. When normalized to the angular areas of their bins, their results give a power law $\Sigma_p(\Delta r)$ with $\gamma \sim -1.5$. Because the depth of the LMC is small compared with most of their separation range, the Elmegreen & Efremov result probably requires little correction for projection effects. However all the other results, for local star-forming regions, involve significant projection effects. Assuming that the depth of each region is at least comparable to its extent in the plane of the sky, the values of $\gamma$ should probably be decreased by unity for estimates of the unprojected $\Sigma_p(\Delta r)$ (Limber 1953; see Peebles 1993), as pointed out by Gomez & Lada (1998), giving $\gamma_{3D}$ in the range –1.2 to –1.8. However the correction for projection depends on the assumed spatial distribution of the stars and the depth of the sample, as shown in the recent toy-model simulations of Bate, Clarke, & McCaughrean (1998).

It should also be pointed out that in most cases deeply embedded pre-main sequence stars are not included in the samples, and a very deep near-infrared census will be required before the pair correlation results can be considered definitive. The effects of extinction on the optically-visible stars and of migration of stars from their birthsites, discussed by NTHN, are also likely to be significant.

No *physical* interpretation of the power-law clustering behavior observed in these regions has been suggested. Larson (1995), Elmegreen & Efremov (1996), and Gomez & Lada (1998) speculate that the power-law reflects the hierarchical distribution of the gas from which the stars form, perhaps due to self-gravity or turbulence. A problem with this interpretation is that the range of power law indices of the correlation function is at variance with the very similar perimeter-area fractal dimensions found for a number of local regions. NTHN argue that hierarchical clustering cannot account for the broad nearest-neightbor distributions (broader than a single Poisson distribution), and instead suggest that a range of cluster sizes is responsible. Their argument is based on the idea that the nearest neighbor distribution of a hierarchy will only reflect the smallest scale of the hierarchy which they assume consists of randomly positioned stars and is identical everywhere. However the nearest neighbor distribution must also include stars formed in larger scales of the hierarchy, or which have migrated from their birth sites, which will broaden the nearest-neighbor distribution over a single Poisson distribution. Non-identical properties of the smallest scales would have a similar effect.

Two papers have independently studied a number of effects that are important for interpreting the correlation function, using "toy models" with various assumed geometries. Houlahan & Scalo (1990) were primarily concerned with applications to continuous structure, but many of the models examined used point source distributions. They illustrated analytically the potentially severe effects of finite spatial extent of the sample (edge effects) and the relative contributions of internal density gradients within individual clusters, a size distribution of clusters, and the cluster-cluster correlations to the correlation function. They also used an analytical model to examine the possibility of signatures of nested (hierarchical) structure. Bate, Clarke, & McCaughrean (1998) have recently presented a careful discussion of many effects relevant to the observational clustering studies listed above, using simulations of toy models. Their work includes a comparison of different estimators of the correlation function and their ability to reduce the edge effects, the dependence of the binary-clustering transition on a number of factors, projection effects due to depth of field, the effects of different spatial distributions (e.g. a single cluster with an internal density gradient, randomly distributed clusters and fractal distributions), and the erasure of structure due to random stellar motions. Bate et al. also present a detailed comparison of the toy models with the observaitonal data, illustrating, among other things, the need for complete surveys over large areas and the fact that a power law correlation function does not necessarily imply an underlying fractal distribution of stars.

In contrast to geometrical toy model interpretations, the question of most interest to us is: *How do the physical processes involved in star formation give rise to the observed power-law clustering properties?* In the present paper we show by numerical hydrodynamical simulations that such power laws can arise in a model in which stellar winds drive expanding shells that are subjected to nonlinear advection and interactions with other shells, and in which gravity acts only in setting a threshold shell column density for instability and star formation. We show how the value of the power law slope $\gamma$ is affected by the age of the stellar sample, the average star formation rate, and by stellar motions after their birth.

## 2. SIMULATIONS

The simulations are in two dimensions, and follow the evolution of a system of interacting wind-driven shells which are subject to nonlinear fluid advection. (The "shells" are actually "rings" in the two-dimensional simulations, but we will continue to use the term "shells" because that would be their initial form in three dimensions.) The system obeys global mass and momentum conservation. The calculations solve hydrodynamical equations describing a highly compressible isobaric fluid that is "forced" or "pumped" at small scales by threshold excitation (star formation here). Such a fluid can be viewed as one in which advection and the corresponding "ram pressure" completely dominate the thermal pressure (Mach number very large), or in which the effective adiabatic index $\gamma$ is near zero (as might approximately apply to the ISM because of the nature of the radiative cooling curve; see Vazquez-Semadeni, Passot, & Poquet 1996; Scalo et al. 1998). Passot & Vazquez-Semadeni (1998) show that these two limits are not equivalent. In this case there is no energy equation to solve; the interactions of fluid elements are completely inelastic. Self-gravity and magnetic fields are neglected, except that local self-gravity is artificially introduced in the form of a threshold instability criterion for the shells. Newly-formed stars are assumed to inject momentum as a wind with a specified kinetic energy. A circularly-symmetric outflow with constant momentum (see, for example, Leitherer 1993 for cluster winds) is injected locally whenever a new star forms at that site, and is assumed to last for a specified time, $\tau_w$. We also allow for a delay time $\tau_d$ between the onset of star formation and the initiation of the momentum input. Star formation is assumed to occur at a threshold column density corresponding to the gravitational instability criterion for an expanding shell, assuming that the critical growth rate at the threshold is a constant. The linear perturbation analysis was generalized to include accretion and local dilatational shell stretching, but these effects turned out not to be important for these sim-ulations. A filament-finding algorithm was used to identify shells, and the column density through the shell was calculated for every computational cell that was part of a shell. Details of the simulations and the filament-finding algortihm are given in a separate paper (Chappell & Scalo 1998, hereafter CS).

The equations describing the evolution of the system are then

$$\frac{\partial \rho}{\partial t} + \nabla \cdot (\rho \mathbf{v}) = 0 \quad (1)$$

$$\frac{\partial \rho \mathbf{v}}{\partial t} + \nabla \cdot (\rho \mathbf{v}\mathbf{v}) = \int\int \frac{\mathbf{x}' - \mathbf{x}}{|\mathbf{x} - \mathbf{x}|} \frac{p_w N_*(\mathbf{x}', t)}{\tau_w} \, d\mathbf{x}' \quad (2)$$

where $\rho$ is the gas surface density, $p_w$ is the total momentum input per massive star, $N_*(\mathbf{x}, t)$ is the number of stars injecting momentum, and $\tau_w$ is the duration of the momentum injection. In the calculations presented here the integration is simply a sum over the eight nearest neighbors. The advection terms are differenced according to a variant of a Van Leer (1977) scheme which is not fully second-order accurate, but is an improvement over a strictly first order scheme with little computational overhead. Modifications were made to minimize anomalous anisotropic effects associated with the numerical viscosity in the scheme, which would otherwise introduce artificial density and velocity fluctuations in expanding shells. Details are given in CS. The boundary conditions were doubly-periodic. The initial conditions consisted of a uniform density field and a Gaussian velocity field with prescribed power spectrum. We examined the effects of varying the initial power spectrum and the resolution ($128^2$, $256^2$, and $512^2$). The scales were normalized such that the lattice spacing was 7.8 pc, so these resolutions correspond to a total region size L of 1, 2, and 4 kpc, but we expect the essential results to apply to smaller or larger scales if the time and velocity scaling is adjusted. The duration of wind momentum injection was taken as $\tau_w = 10^7$ yr, corresponding to cluster winds; scaling to smaller size systems would reduce $\tau_w$ to about $10^6$ yr for a system size of a few parsecs, although it should be noted that the protostellar winds at these scales are not spherically symmetric as in the simulation model.

A series of $256^2$ simulations with a flat initial energy spectrum is presented here. The simulations were integrated for about 2 Gyr (for the adopted length scale), long enough for initial transients to disappear and to study the temporal evolution of the system. We point out that these very long integrations were made possible by neglect of physical processes beside advection, and by the adoption of a first-order difference scheme. A detailed presentation of the models, including the effects of variations of parameters and initial conditions, is given in CS.

All the simulations evolve into a network of irregularly shaped filaments (in two dimensions) which cover a large range of sizes and which are the products of the distortion of the originally symmetric star-forming shells by interactions with other shells and by advection (the distinction is not clear-cut, since most of the mass ends up in the filaments). Sometimes filament interactions lead to nearly spherical "clumps" which may or may not be dense enough to form a star. Often an expanding shell produced by one "star" or "cluster" compresses gas along the filament in which it was born, stimulating further star formation and sometimes resulting in groups or chains of stars The overall filamentary structure is not dependent on the existence of the wind energy input, but is an inevitable result of advection and the high compressibility brought about by the absence of pressure. Simulations without stellar forcing (presented in CS; see also Scalo et al. 1998) develop similar structure, although of course with no energy input the structure is eventually concentrated on large scales, and the velocities monotonically decrease with time.

Examples of the density field (left) and the spatial distribution of stars of various maximum ages (right) are shown in Fig. 1. The three cases shown correspond to different average star formation rates, which are controlled by the assumed threshold shell column density required for gravitational instability and star formation. The filaments are all thin because there is no pressure; their thickness is set by numerical diffusion. However analogous simulations that include pressure (and cooling and self-gravity) develop a similar appearance (see the examples in Vazquez-Semadeni, Passot, and Poquet 1996 and references therein; also Scalo et al. 1998).

## 3. Results

The density of companions as a function of separation is computed here in the same way as used by Larson (1995) and others. We define a number of separation bins $\Delta r_j$ each of which has an associated area $A_j = \pi(\Delta r_{j+1/2}^2 - \Delta r_{j-1/2}^2)$. For each object $i$, the companion density at separation $r_j$ is

$$\rho_i(\Delta r_j) = n_{ij}/A_i \quad (3)$$

where $n_{ij}$ is the number of objects in the j-th separation bin for the i-th star. The average pair density function is then an average over all the stars

$$\rho_p(\Delta r_j) = \frac{1}{N} \sum_{i=1}^{N} \rho_i(\Delta r_j) \qquad (4)$$

where $N$ is the total number of objects. We use the notation $\rho_p$ rather than $\Sigma_p$ for the simulations because, although they take place in two dimensions, there is no projection effect.

A discussion of different techniques to evaluate the average pair density function in order to correct for edge effects is given by Bate et al. (1998). However edge effects are minimal in the present simulations because the simulation domain is much larger than the clustering scale and because of the periodic boundary conditions.

The positions of stars in the simulations (and in real star-forming regions) reflect both the spatial distribution of gas out of which they formed, which varies with time, and the motion of the stars since the time of their birth. For this reason we studied $\rho_p(\Delta r)$ using the positions at which the stars were formed in the past (reflecting the gas distribution) and also according to the present positions computed assuming that each star has moved ballistically since its birth at the velocity of the gas from which it formed. The latter procedure neglects any gravitational scatterings which the stars experience, but should give some idea of the effects of stellar motions in smearing the separation distribution.

In addition, we computed $\rho_p(\Delta r)$ for stars of different maximum ages $T_{max}$. This age effect is important in the present models even for the case in which stellar motions are neglected, because a larger range in stellar ages reflects a larger range of past configurations of the gas from which the stars formed. For example, a large value of $T_{max}$ will contain stars formed from all gas spatial distributions in a given region which existed from the present time back to a time $T_{max}$ in the past when the gas spatial distribution may have been very different. This averaging over past gas distributions has the effect of broadening the distribution of separations. In order to make the discussion general, we give times in units of the size $L$ of the computational domain divided by the velocity dispersion $c$ of gas in the region. For the actual simulations the velocity dispersion does not depend much on the parameters (CS), so we take a typical gas velocity dispersion of 5 km/sec. With $L = 1$ kpc, the time unit is $L/c = 2 \times 10^8$ yr. However the results should be general enough that they should apply to any star-forming region if times are measured in units of $L/c$. For example, for pre-main sequence stars in a region of size a few parsecs, the time unit may be of order $10^6$ yr.[1]

Our results are displayed in Figure 2. Each correlation function was computed by summing the number of pairs in each separation bin over a large number of "snapshots" (like those in Fig. 1) sampled over the last half of each simulation run. The three panels correspond to three different average star formation rates, as in Fig. 1, controlled in the simulations by the assumed threshold shell column density required for gravitational instability. Within each panel the resulting $\rho_p(\Delta r)$ is displayed for maximum ages ranging from $T_{max} = 0.025$ (bottom) to 0.25 (top). The filled circles correspond to $\rho_p(\Delta r)$ computed from the initial birth sites, while the open circles show the effect of allowing for stellar migration. For both cases the logarithmic slopes of a power law fit to the section that appears to be a power law are indicated. We tested for effects of numerical resolution by computing $\rho_p(\Delta r)$ for several models with $512^2$ resolution. The fitted slopes $\gamma$ were identical to those obtained for the $256^2$ simulations within $\pm 0.1$. We also allowed the SFR to vary by changing the time delay between onset of instability and the beginning of wind momentum input, rather than varying the instability threshold but, within the range of SFRs available using this approach, the results were very similar. Because the simulations use boundary conditions that are doubly-periodic, the flattening at large separations simply reflects the lack of correlation between widely-separated regions. In contrast, the observed correlation func-tions for local star-forming regions exhibit a sharp dropoff at large separations because they refer to a localized enhancement of young stars, and so there are simply very few pairs of stars at separations corresponding to the observed region size. For this reason the behavior of the simulations and observations at large separations approaching the size of the region are not comparable.

---

[1] As pointed out above, protostellar winds at these scales are not spherically symmetric. Although we do not know in detail what the effect of such "incomplete shells" would be, we suspect that the correlation function would not be affected much unless the outflows were extremely well collimated, since the advective gas structures built from the small-scale input is essentially independent of the input geometry. However because the shells would have smaller areas (lengths in the simulations), the characteristic shell interaction time would be somewhat larger, implying that the system will be "less evolved" for a given L/c. However it should be remembered that the simulations were run for very long times, so transient evolutionary effects like this are not significant. However the SFR corresponding to a given L/c may be different than given for the simulations, to the extent that the SFR depends on the rate of shell interactions.

Several trends are immediately obvious from Fig. 2. First, $\rho_p(\Delta r)$ is flatter for larger star formation rates. This occurs because the "clusters" become larger at larger SFRs, or equivalently, because different small-scale clusters become more spatially correlated, i.e. higher-order clustering becomes more important. There is no real distinction because the clustering in the power law regime is scale free, and the higher SFR runs simply result in more power at larger scales (see Fig. 1 and imagine averages taken over many such snapshots).

Second, increasing the maximum stellar age, $T_{max}$, averages over a larger range of past gas spatial distributions as explained earlier, and so also flattens $\rho_p(\Delta r)$.

Third, as expected, inclusion of stellar migration flattens the correlation function for separations less than the average distance which stars have had time to move. This effect is important for dimensionless times greater than about 0.1 on small scales, giving support to the claim of NTHN that the effect of stellar motions is important. The slope of the correlation function remains finite for separations affected by stellar motions because, in the simulations, stars are continually being formed. Thus the youngest stars at small separations still reflect correlations due to the gas from which they formed. In contrast, the toy model simulations of Bat et al. (1998) assume that all the stars are formed simultaneously, so that stellar motions eventually erase all the structure, resulting in a flat correlation function at small scales. In principle this difference could be used to distinguish between ongoing star formation and star formation that has ceased sometime in the past, but in practice, at least for bound clusters, the internal density profile will continue to give a finite slope to the small-scale correlation function even long after star formation has ceased (Bate 1998, personal communication).

If we interpret our 2-dimensional simulation results for $\rho_p(\Delta r)$ in terms of the *unprojected* observed functions, our logarithmic power law slopes cover the observed values, –1.2 to –1.8. It is questionable whether our results should be compared to observed regions in which star formation has terminated (e.g. the Cepheids in the LMC), since our model allows star formation to proceed up to the present time. Certainly our results suggest that one should not expect any universal value for the power-law index of $\rho_p(\Delta r)$, which should depend on age of region and star formation rate.

## 4. Discussion

Our simulations can easily account for the range in pair correlation slopes found in the observed samples, assuming that the unprojected simulation slopes would be similar if they were carried out in three dimensions instead of two, and that the projection of the true pair surface density function to the observed function will flatten the power law index by unity. More realistically, the correction for deprojection will depend on the spatial distribution of stars and the depth, as discussed by Bate et al. (1998). In three dimensions the evolution of our simulations would be dominated by interacting advecting shells, rather than filaments, but we have no reason for thinking that opening up a third dimension for these processes should result in any significant changes for the scales or slopes of the correlation function.

A comparison of our predicted dependence of $\gamma$ on maximum stellar age and star formation rate with observations is currently problematic, for a number of reasons which we now discuss. For the local star-forming regions the stellar ages are very uncertain because they depend on various empirical calibration uncertainties and on the adopted theoretical evolutionary tracks from which the ages are estimated. An instructive example of the problems can be found in the comprehensive study of optical pre-main sequence stars in the Orion Nebula Cluster by Hillenbrand (1997). Similarly, the average star formation rate cannot be estimated without knowledge of the age of a region. Of all the local regions, only Ori OB (NTHN) and $\lambda$ Ori (Gomez & Lada 1998) are clearly older than the rest. Ori OB does have the flattest slope (-0.15) of all the regions, but $\lambda$ Ori has a steeper slope (-0.49). Unfortunately there is no way to estimate the star formation rates that occurred when the regions were actively forming stars, which should also affect the slope. In addition, these regions cannot be properly compared with the models because star formation has essentially terminated in the observed regions, but continues to the present in the models.

For the younger regions, one might assume they all have about the same age and search for the predicted correlation between steeper slopes and smaller SFRs by comparing the mean nearest neighbor separation or the surface density of stars. Such a comparison cannot be carried out with the available data. Concerning the first suggestion, the mean nearest neighbor separations given by NTHN do not simply reflect the mean stellar separation and surface density, but must also strongly reflect the degree to which the stars are clustered, as does the slope $\gamma$. These two quantities will naturally be (anti-) correlated because they are measuring the same effect. Notice that this effect works in the opposite direction from the model predictions.

One could try to use the number of survey stars and area surveyed to estimate the surface density and hence SFR, but a number of effects preclude this. First, and probably most importantly, as pointed out by NTHN, the samples differ greatly in depth, bias, incompleteness, and contamination, even for regions at about the same distance. Second, extinction may severely affect the derived values of $\gamma$ (NTHN); note that except for $\rho$ Oph, only the optically-visible stars and/or X-ray sources were included in the samples. Third, the derived slope and surface density depend on the areal coverage with respect to the most tightly clustered stars. For example, imagine a tightly clustered group of stars surrounded by a loosely-clustered population. If the survey only covers the tight cluster, one may obtain a steep correlation function (simply reflecting the radial surface density profile within the cluster) and a large surface density, while if a larger area were surveyed, one would obtain a flatter correlation function (since the correlation function basically represents the frequency distribution of separations for point sources; see Houlahan & Scalo 1990) and a smaller surface density. This sample selection effect thus works in the opposite direction to the predicted correlation of $\gamma$ with SFR. Our examination of the data suggests that this is a large effect. For example, the Cham I cluster was apparently only surveyed over a small area ($\sim 1$ deg$^2$) and gives a rather steep $\gamma = -0.57$ but a very large surface density, while $\rho$ Oph has been surveyed well outside the deeply embedded cluster ($\sim 40$ deg$^2$) and yields a flatter slope $\gamma = -0.36$ and a much smaller surface density. This areal coverage effect could work in the opposite direction. If the survey was too concentrated on a cluster, the motions of the stars would randomize the correlation function, counteracting the effect of the radial gradient in mean cluster surface density.

It seems clear that a very carefully-designed homogeneous set of observations, including deep near-infrared counts of several regions to the same limiting absolute magnitude, and age determinations based on recent theoretical evolutionary tracks and conversion calibrations, will be necessary in order to test the dependence of slope on age and SFR predicted by the present models. Some such data do exist for a few embedded clusters (e.g. Luhman & Rieke 1998, Williams et al. 1995, 1996), but these clusters have only been studied with respect to the form of the initial mass function at small masses. A difficult problem is that, for comparison with the present simulations, which typically include several star-forming clusters, the data would have to be combined with data for larger surrounding areas that have the same limiting bolometric absolute magnitude as the near-infrared embedded cluster surveys. The DENIS (Epchtein 1997) and 2MASS (Skrutskie et al. 1997) surveys should improve the situation, although a careful treatment of corrections for depth effects and filtering to remove large-scale structure in the star density will be required in order to compare with the present models.

Another prediction of our simulations (which would probably be common to other models as well) is that if the region is old enough so that stars have had time to move significantly from their birthsites, $\rho_p(\Delta r)$ should exhibit a relatively sharp flattening at scales smaller than the average distance over which stars have moved. This effect is best seen for the intermediate star formation rate case in Fig. 2. The flattening can be relatively large, with a change of slope of 0.3 to 0.5 for the models we have studied (and presumably larger if older stars were included). Although other physical effects might play a role, it seems significant that Gomez & Lada (1998), in their study of Orion stars, found a clear change in slope in the clustering regime at an angular separation of about 0.5°, which corresponds to about 4 pc for Orion. This scale is reasonable for the stellar spreading effect if the stellar velocity dispersion is 1–3 km sec$^{-1}$. Some evidence for a similar effect can be seen in the data of NTHN for Ori A ($\sim 0.8°$), $\rho$ Oph ($\sim 1°$), and especially Lupus ($\sim 0.1°$). No such break would be seen in a region which is either extremely young (or the stellar velocity dispersion very small), in which case the "stellar migration break" may be within the binary regime, or relatively old, in which case the spreading will have affected even the largest observed separations. The Elmegreen & Efremov (1997) result for the Cepheids in the LMC might seem like a problem in this regard, since the maximum age of the sample ($2-3\times 10^7$ yr) is larger than the local samples discussed above, suggesting that the spreading effect should have reached large scales, while the correlation slope inferred from their plot is relatively steep (–1.5, with no need for deprojection). However it must be remembered that the absolute scale of separations they studied (0.1–2 kpc) was much larger than for the local regions. Even with a velocity dispersion of 5 km sec$^{-1}$, stars of this age would have only travelled about 150 pc, which is near the lower limit of their separation range.

It is ironic that even though we can provide a physical mechanism for the power-law correlations, the models do not clearly provide evidence for one of the geometrical hypotheses suggested by others for the correlations: that they are due to a nested hierarchy of the gas from which the stars formed, or that they reflect a power law distribution of cluster sizes. Both effects are evident in the simulations, and they are not even clearly delineated, suggesting that these ways of conceptualizing the structure are in some ways too simple or reductive. A discussion of the ambiguity inherent in using the correlation function to distinguish between a fractal distribution, a random distribution of clusters, or a surface density gradient within individual clusters is given in Bate et al. (1998).

Our main result is to demonstrate that a power law form of the correlation function, or pair density separation function, can result from a model in which star formation occurs as a threshold phenomenon in a field of shells driven by stellar winds, advection, and shell interactions. Except for the effects of advection, the model is phenomenologically a close cousin of the Norman & Silk (1980) analytical model for wind-driven support and star formation in dense molecular clouds. We feel that an explanation of the observed power laws in terms of physical processes, rather than ascribing them to some unknown process that produces fractals or hierarchies, is a major advance in understanding the phenomenon. We point out that the perimeter-area fractal dimensions we have computed for a few of our simulations are close to 1.1 (since the density field is basically a collection of filaments with a large range in sizes), yet we find a variety of slopes $\gamma$ for the pair correlation function, because the correlation function of the stars is reflecting more than just the instantaneous dimension of the gas from which they formed, although that certainly plays a role. That a power law correlation function need not reflect a fractal spatial distribution has also been emphasized by Bate et al. (1998).

We suspect that other types of physical models will also be able to account for at least the range of slopes of the correlations, since the correlation function is not a very sensitive indicator of either structure or physical processes. A discussion of the nearest-neighbor distribution function (which is only a one-point statistic, but at least is not a moment of a distribution, as is the correlation function) and comparison with the distributions presented in NTHN is postponed to a separate paper. For the present we have at least shown that the physical model adopted here is capable of accounting for the power law slopes, and have predicted how the slope should vary with the star formation rate and the maximum stellar age for our model, although much more observational work is required before an observational test is warranted. We also found that stellar motions may introduce a "migration break" in the correlation function at relatively large scales, and suggest that this effect may explain the changes in correlation function slopes that are seen in some regions at large separations.

Finally, we want to emphasize that, while the study of the observed clustering properties of young stars is undoubtedly important in understanding star formation, the correlation function itself may not be well-suited to provide physically interesting information. The correlation function is useful in studying cosmological large-scale structure because the the universe is (approximately) statistically homogeneous on the scales of the surveys, and includes many groupings of galaxies over a range of scales, which then can give a statistically meaningful correlation function. The same is approximately true of our simulations, because the simulated region is much larger than most of the clustering and because we have averaged over a large number of snapshots, both of which establish a statistically homogeneous sample. However if the surveyed region is not statistically homogeneous (i.e. contains structure on the scale of the survey or contains only a few clusters), which is true for most of the local star-forming regions, the correlation function only yields information that can be more easily estimated by other means. For example, if a region only contains several clusters, a power law portion of the correlation function will only reflect a mixture of the surface density gradients within each cluster and the size distribution of the clusters, and the correlation function will exhibit recorrelation "bumps" at larger scales corresponding to the separations of the clusters. (Examples are given in Houlahan and Scalo 1990. See Miesch and Bally 1994 for a careful attempt to use filtering to eliminate statistical inhomogeneity in estimating velocity correlation functions of interstellar structures, as well as a discussion of the dangers.) This is basically what we referred to earlier as the "areal coverage" problem. For this reason we urge observers to explore other measures of the spatial distribution of young stars, especially those explicitly taking into account the ages and masses of stars, quantities that can in principle be measured and which will provide strong constraints on models.

We thank the referee, Mathew Bate, for constructive comments and suggestions. This work was supported by NASA Grant NAG 5-3107.

FIGURE CAPTIONS

Figure 1. Typical snapshots of the logarithm of the density field of the simulations (left panels), and positions of stars of various maximum ages (right panels), for three different star formation rates (top to bottom). For the stars, filled circles, open circles, and crosses correspond to maximum ages of 0.025, 0.125, and 0.25 in units of size of region divided by velocity dispersion of gas in filaments, L/c.

Figure 2. Stellar pair correlation functions averaged over many samples of each simulation, for the three star formation rates (left to right) corresponding to Figure 1. For each star formation rate, the correlation functions are shown for different maximum stellar ages of the sample. In dimensionless units of L/c these times are (from bottom to top) 0.025, 0.05, 0.125, and 0.25. Filled circles are for stars at their birthsites, open circles include stellar motion at the velocity of the local gas from which they formed. Logarithmic slopes of least squares power law fits to the results are indicated.

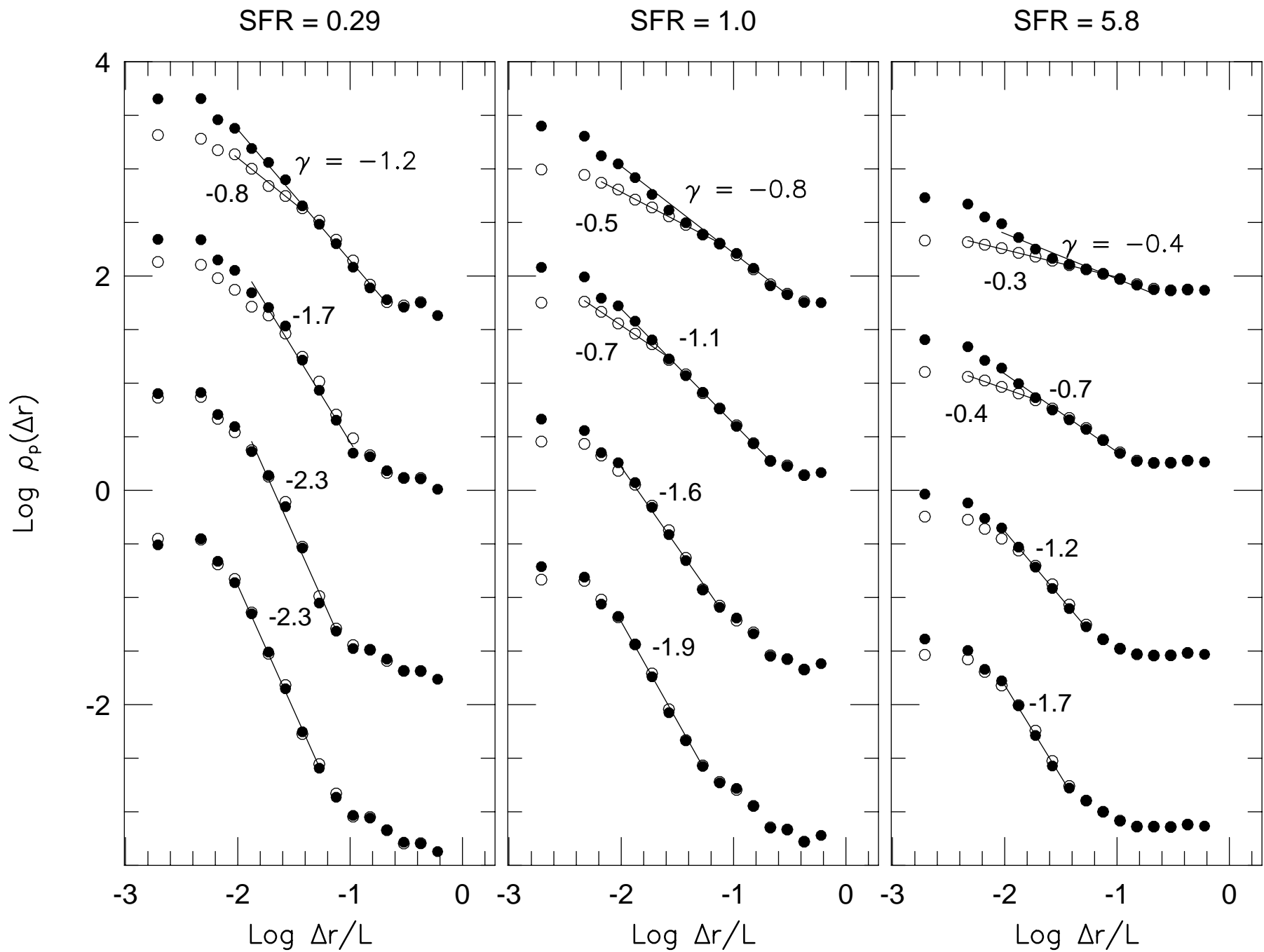

## SFR = 0.29

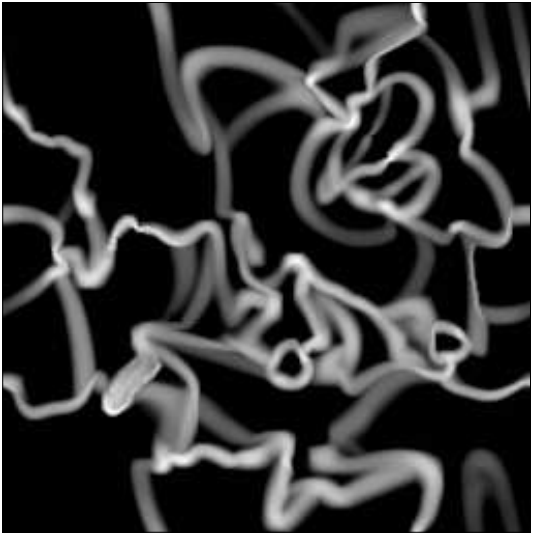 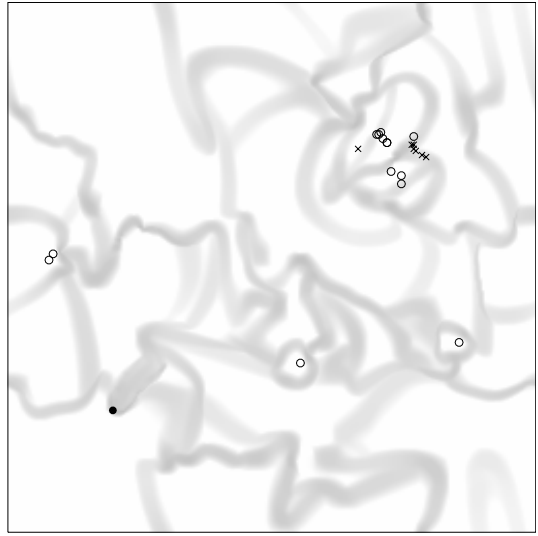

## SFR = 1.0

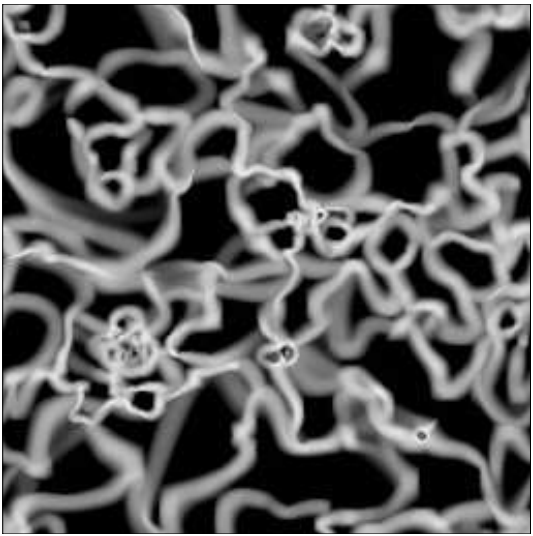 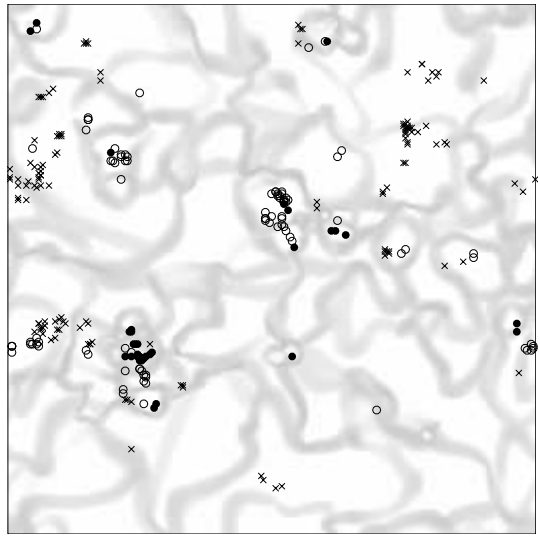

## SFR = 5.8

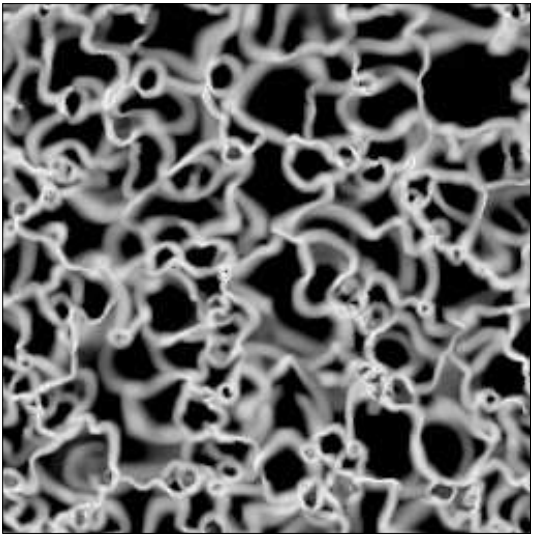 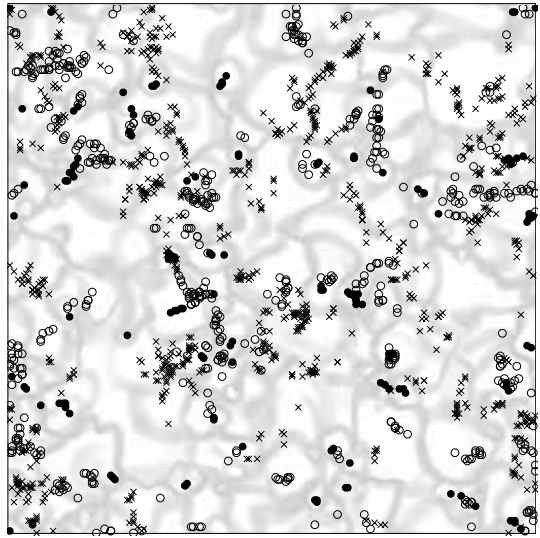